\documentclass[prd,onecolumn,preprint,nofootinbib]{revtex4}
\usepackage[a4paper,margin=2.4cm]{geometry}
\usepackage{natbib}
\usepackage{graphicx}
\usepackage{amsmath}
\usepackage{booktabs}
\usepackage[colorlinks=true,linkcolor=blue,citecolor=blue,urlcolor=blue]{hyperref}
\graphicspath{{./}{figures/}}

\begin{document}
\title{Which is the best day of the week to   submit to arXiv:astro-ph?}

\author{Chirag Sharma}
\altaffiliation{E-mail : csharma@kgpian.iitkgp.ac.in}
\author{Shantanu Desai}
\altaffiliation{E-mail: shntn05@gmail.com}

\affiliation{$^{1}$Department of Physics, Indian Institute of Technology Kharagpur, Kharagpur 721302, India}

\affiliation{$^{2}$Department of Physics, Indian Institute of Technology Hyderabad, Kandi, Telangana 502284, India}

%\date{Draft: \today}

\begin{abstract}
We examine whether the citations received by refereed astronomy papers vary systematically with the day of the week of their arXiv preprint release. We use NASA ADS citation data for all refereed articles published in 2020--2023 in six journals (MNRAS, ApJ, ApJL, ApJS, A\&A and JCAP; 37,173 papers with astro-ph primary preprints), combined with version-1 submission timestamps from the arXiv API. We group papers both by submission weekday and by the day their announcement batch appears on arXiv, accounting for arXiv's weekday deadline and lack of weekend announcements. Given the heavy skew of citation distributions, our primary statistic is the median citations per group, with bootstrap confidence intervals. Papers submitted on weekends, all of which appear in the Monday listing, have median citations significantly lower than weekday submissions in five of the six journals, by 12--26 \%; the sixth (JCAP) shows the same direction, but not significantly, consistent with its small weekend sample. Median citations are indistinguishable across the five weekdays. These results extend the recently reported weekend citation disadvantage in particle and nuclear physics to astronomy. The link is correlational, consistent with either compositional variations in weekend submissions or reduced visibility in the enlarged Monday listings.
\end{abstract}
\maketitle

%\medskip
%\noindent PACS: 01.30.-y % TODO: Prof. Desai to confirm/extend (01.30.-y = physics literature and publications)

\section{Introduction}
\label{sec:intro}

Citation counts are widely used as a proxy for the impact of scientific work, influencing hiring, funding, and career trajectories. Ideally, citations should track scientific merit alone. The arXiv preprint server, however, also acts as an attention mechanism: a large fraction of the astronomy community scans the daily astro-ph listing, and a paper's early visibility in that listing plausibly influences who reads, remembers, and ultimately cites it.

Several studies have documented visibility effects on arXiv. Dietrich~\cite{Dietrich2008} showed that papers appearing near the top of the daily astro-ph listing receive roughly twice the citations of papers listed lower, an effect driven in part by authors who deliberately submit  immediately after the daily deadline to secure top positions. Haque and Ginsparg~\cite{HaqueGinsparg2009} confirmed and refined this result, finding that articles in the first announcement position received a substantial median citation boost even when the position was obtained accidentally rather than through deliberate timing --- direct evidence that visibility alone, independent of paper quality, leaves long-term traces in the citation record. A follow-up study~\cite{HaqueGinsparg2010} identified additional positional and submission-timing effects. Day-of-week patterns in journal submission and acceptance have also been examined outside physics~\cite{Boja2018}.

Most directly relevant to the present work, Delorme et al.~\cite{Delorme2025} analyzed
328{,}936 manuscripts in the hep and nucl arXiv categories submitted between January 1991
and March 2025, binned by submission day in New York local time. They found the average
citation count essentially uniform across weekday submissions (mean $\approx 40.4$
citations, with a day-to-day scatter of about 1\%), explicitly testing and rejecting the
folklore that Wednesday submissions are better cited. Weekend submissions averaged 33.2
(Saturday) and 34.2 (Sunday) citations, about 6--7 fewer than the weekday mean and more
than $10\sigma$ below it.
% TODO (Chirag): from their Fig. 2(b), add their M-F versus weekend mean-citation
% values here in their own units. Do NOT reuse numbers from our earlier notes.

To our knowledge, no comparable analysis has been performed for Astronomy (astro-ph) using only refereed journal publications. This paper fills that gap. We analyze six top journals spanning a range of sizes, publishers, and geographic author bases, and we pay particular attention to a structural feature of arXiv that any day-of-week analysis must respect: arXiv batches submissions at a weekday 14:00~ET deadline and makes no announcements on weekends, so that every submission made between Friday 14:00~ET and Monday 14:00~ET appears together in Monday's listing~\footnote{\url{https://info.arxiv.org/help/availability.html}}. ``Posting day'' is therefore two distinct variables --- the day the author submitted, and the day the paper appeared --- and we analyze both.

Our methodology follows the median-based approach of Rithvik and Desai~\cite{Desai2025,Rithvik2}, motivated by the extreme right skew of citation distributions~\cite{Seglen1992}, for which the mean is dominated by a small number of highly cited papers.

\section{Data and methodology}
\label{sec:data}

\subsection{Sample construction}
For our analysis, we considered the journals MNRAS, ApJ, A\&A, ApJL, JCAP, and ApJS. These are some of the top astrophysics journals in terms of impact factors, which also publish a large number of papers every year~\cite{Desai2025,Rithvik2}.
For each journal we queried the NASA Astrophysics Data System (ADS)~\cite{ADS} for all refereed articles published between January 2020 and December 2023, retrieving bibcodes, identifiers, and current citation counts. For ApJ, we use the main journal only; ApJL and ApJS are analyzed separately. From each record we extracted the arXiv identifier where present; records without an arXiv preprint were excluded. For every arXiv identifier we then retrieved the version-1 (v1) submission timestamp and the primary category from the arXiv API, and restricted the sample to papers whose \emph{primary} category is in astro-ph, since the primary category determines the listing in which a paper appears. Table~\ref{tab:funnel} summarizes the sample construction.

\begin{table}[ht]
\centering
\caption{Sample construction. ``No arXiv'' is the fraction of refereed articles without a retrievable arXiv identifier; ``non-astro-ph'' counts preprints whose primary category lies outside astro-ph.}
\label{tab:funnel}
\begin{tabular}{|l|r|r|r|r|}
\toprule
\hline
\textbf{Journal} & \textbf{ADS records} & \textbf{No arXiv} & \textbf{Non-astro-ph} & \textbf{Final sample} \\
\hline
\midrule
MNRAS & 15{,}856 & 1{,}250 (7.9\%) & 268 & 14{,}338 \\
ApJ   & 12{,}107 & 1{,}604 (13.2\%) & 350 & 10{,}153 \\
A\&A  & 8{,}754  & 775 (8.9\%) & 92 & 7{,}886 \\
ApJL  & 2{,}697  & 290 (10.8\%) & 82 & 2{,}325 \\
JCAP  & 2{,}983  & 24 (0.8\%) & 1{,}381 & 1{,}578 \\
ApJS  & 1{,}184  & 231 (19.5\%) & 60 & 893 \\
\bottomrule
\hline
\end{tabular}
\end{table}

Two features of Table~\ref{tab:funnel} deserve attention. First, the fraction of refereed papers without an arXiv preprint differs across journals (from under 1\% for JCAP to about 20\% for ApJS), so each journal sample is subject to a different, unmodeled selection. Second, nearly half of JCAP articles have a primary category outside astro-ph (predominantly hep-ph, hep-th, and gr-qc), reflecting that journal's scope; only its astro-ph-primary papers enter our analysis.

Because papers are selected by journal publication year but grouped by  arXiv posting date, many preprints in the sample were posted before 2020, reflecting the typical lag between preprint posting and journal publication.

\subsection{Day-of-week assignment}

Each v1 timestamp was converted from UTC to US Eastern time using the IANA timezone database, which correctly handles daylight-saving transitions; a fixed UTC offset would misclassify submissions within an hour of the deadline for half of each year. We then define:
\begin{enumerate}
\item \textbf{Submission day} (seven values): the local weekday at the moment of {\tt v1} submission.
\item \textbf{Announcement day} (five values, Monday--Friday): the day of the announcement batch in which the paper appeared, obtained by rolling any submission after 14:00~ET to the next day and any submission falling on Saturday or Sunday forward to Monday.
\end{enumerate}

We label each announcement batch by the day its submission window closes. Under arXiv's schedule for submissions, the batch closing Monday 14:00~ET --- which contains all weekend submissions --- is announced at 20:00~ET on Monday, 
while the batch closing Friday 14:00~ET is announced on Sunday evening, there being no Friday or Saturday announcements. US holidays occasionally merge or delay batches; we do not model this, and it mis-assigns a small number of papers per year by one day.

The day-assignment logic was validated three ways: unit tests including explicit daylight-saving edge cases; an independent from-scratch re-implementation of the deadline rule, which agreed with the pipeline on all tested edge cases (weekend submissions and submissions within minutes of the deadline); and re-retrieval of the stored timestamps from the live arXiv API, which matched exactly.

\subsection{Statistical methods}
\label{sec:stat}
Citation distributions are strongly right-skewed~\cite{Seglen1992,Redner05}: a single paper with thousands of citations can dominate the mean of a bin, making it less representative of a typical paper. Our primary statistic is therefore the \emph{median} citation count per group, with 95\% percentile-bootstrap confidence intervals ($10^4$ resamples), following ~\cite{Desai2025}. We report  the means alongside for comparison with earlier work. Group differences are assessed with the rank-based Kruskal--Wallis test across day bins and the Mann--Whitney $U$ test~\cite{astroML} for the pre-specified weekend-versus-weekday comparison; pairwise weekday comparisons use Holm correction. As robustness checks we repeat the analysis stratified by publication year (2020--2023 papers have had different times to accrue citations) and with year-normalized citation counts.

\section{Results}
\label{sec:results}

\subsection{Weekend versus weekday submissions}

Table~\ref{tab:weekend} presents the central result. In all six journals, papers submitted on weekends have a lower median citation count than papers submitted on weekdays. The deficit is statistically significant (Mann--Whitney) in five journals; in JCAP the direction is the same but the test does not reach significance, which is unsurprising given its relatively small  weekend sample of 104 papers.

\begin{table}[ht]
\centering
\caption{Weekend versus weekday submissions, by journal. Medians are citation counts as of the data retrieval date (June 2026). $N_{\rm we}$ is the number of weekend-submitted papers.}
\label{tab:weekend}
\begin{tabular}{|l|r|r|r|r|r|r|l|}
\toprule
\hline
\textbf{Journal} & $N$ & $N_{\rm we}$ & Med$_{\rm we}$ & Med$_{\rm wd}$ & Mean$_{\rm we}$ & Mean$_{\rm wd}$ & \textbf{MW} $p$ \\
\hline 
\midrule
MNRAS & 14{,}338 & 983 & 13 & 16 & 21.0 & 26.5 & $3.9\times10^{-13}$ \\
ApJ   & 10{,}153 & 929 & 14 & 18 & 22.4 & 28.4 & $1.3\times10^{-9}$ \\
A\&A  & 7{,}886  & 555 & 15 & 18 & 34.8 & 34.1 & $7.2\times10^{-6}$ \\
ApJL  & 2{,}325  & 218 & 22 & 25 & 34.5 & 49.3 & $0.020$ \\
ApJS  & 893      & 87  & 17 & 23 & 34.9 & 58.1 & $5.7\times10^{-4}$ \\
JCAP  & 1{,}578  & 104 & 15 & 19 & 27.5 & 33.8 & $0.15$ \\
\bottomrule
\hline
\end{tabular}
\end{table}

\begin{figure}[ht]
\centering
\includegraphics[width=\textwidth]{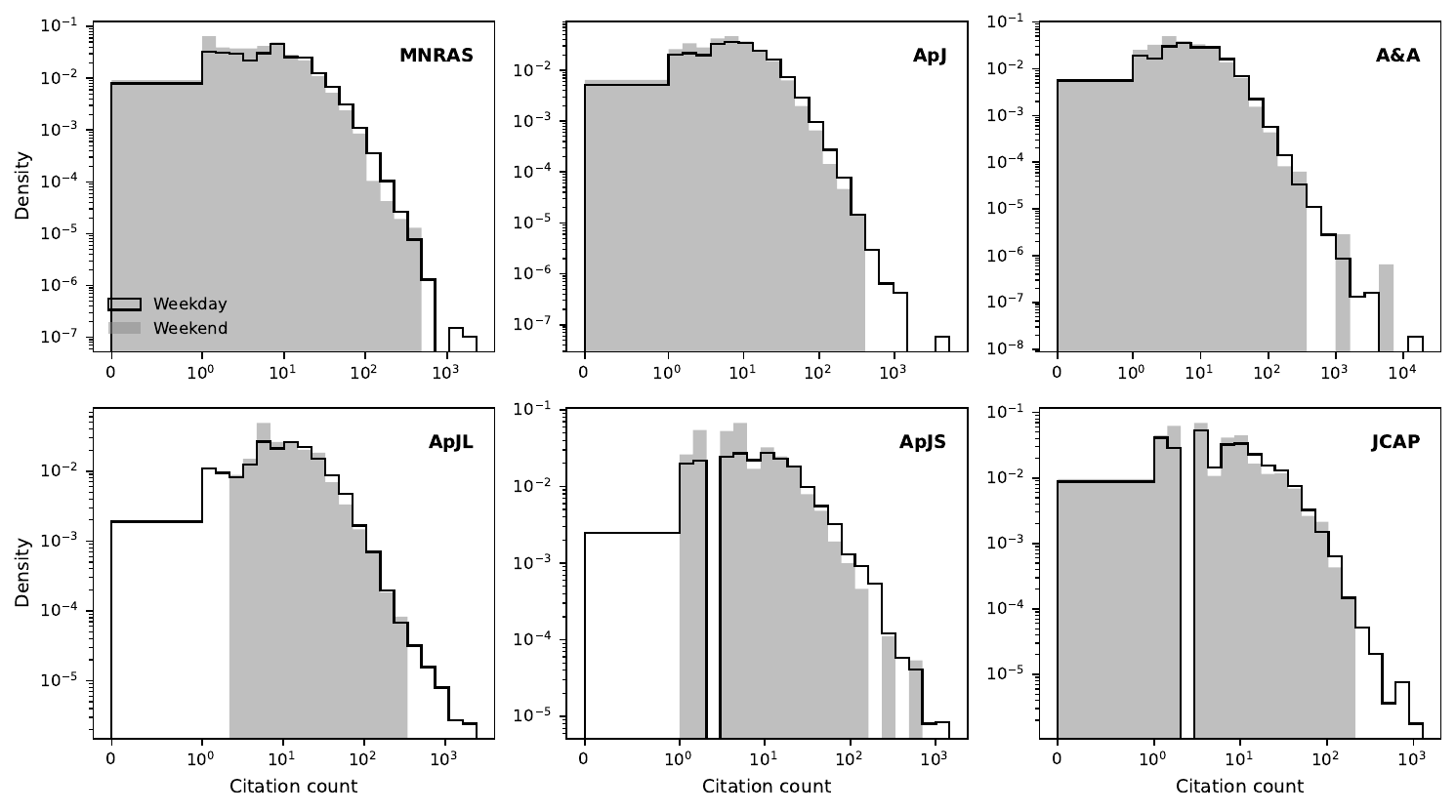}
\caption{Histograms of the number of citations using 20 logarithmically spaced bins, for weekend-submitted (shaded) and weekday-submitted papers, for each of the six journals. The citation distributions follow approximate power laws, motivating the use of median-based statistics.}
\label{fig:hist}
\end{figure}

\begin{figure}[ht]
\centering
\includegraphics[width=\textwidth]{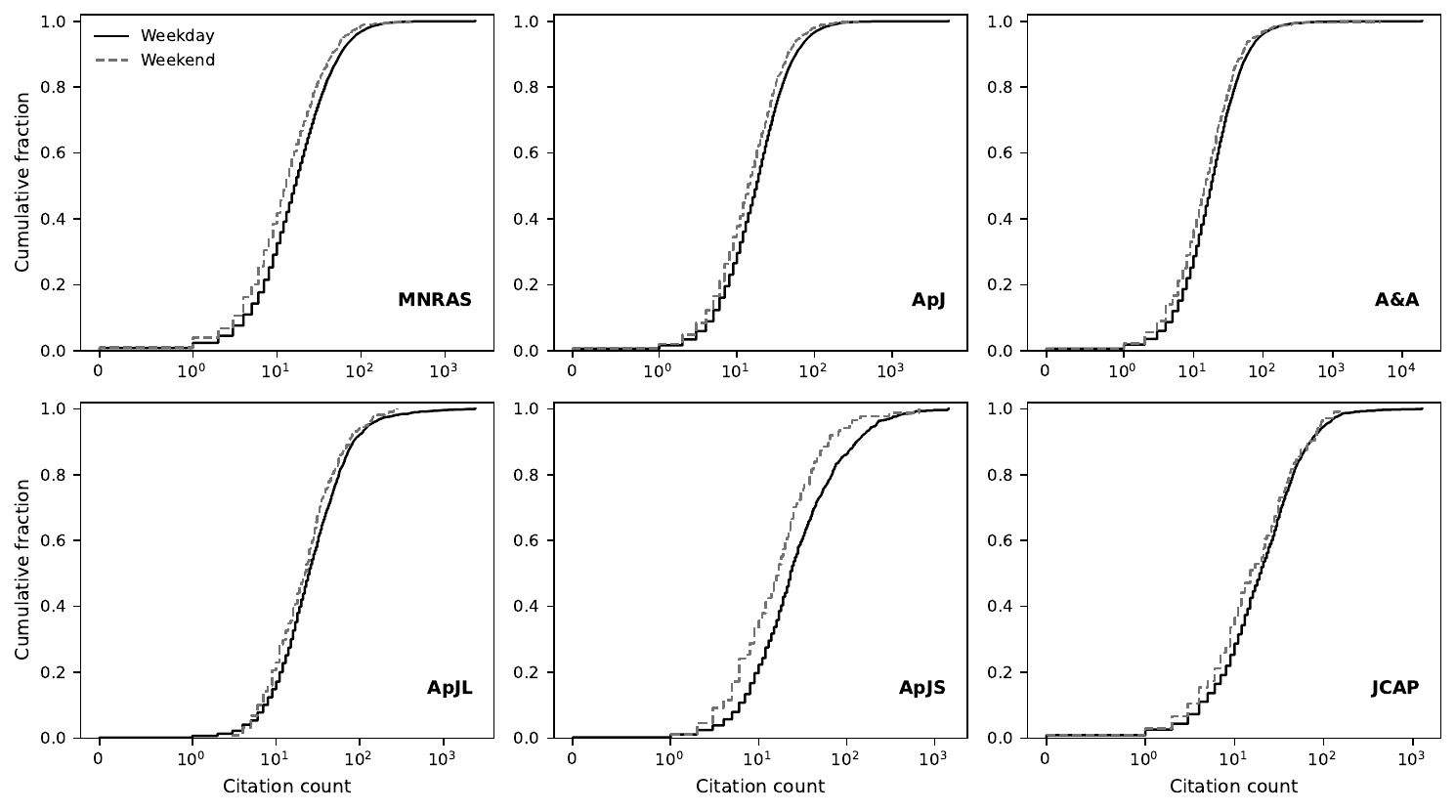}
\caption{Unbinned cumulative distributions of the number of citations for weekend-submitted and weekday-submitted papers, for the same six journals shown in fig.~\ref{fig:hist}.}
\label{fig:cdf}
\end{figure}

In relative terms,  the median deficit ranges from 12\% (ApJL) to 26\% (ApJS). Among the five weekdays themselves, median citations are statistically indistinguishable within each journal (e.g., 16--17 for MNRAS, 17--19 for ApJ). The dominant day-of-week structure in these data is therefore essentially binary: weekend versus weekday. The full citation distributions underlying this comparison are shown in Figs.~\ref{fig:hist} and~\ref{fig:cdf}; the strong right skew visible in these distributions  motivates the use of  median-based statistics in Sect.~\ref{sec:stat}.

The A\&A row illustrates why the median is the appropriate primary statistic here: the \emph{mean} difference in A\&A is slightly positive ($+0.7$ citations), in apparent contradiction with its clearly lower weekend median and its highly significant rank test. Inspection shows this is driven by a small number of highly cited large-collaboration papers that happen to fall in the small weekend bins --- most notably the Gaia Data Release 3 summary paper~\cite{GaiaDR3}, submitted on a Saturday and with 4772 citations at the retrieval epoch, and the Gaia Early Data Release 3 astrometric solution paper~\cite{GaiaEDR3}, submitted on a Sunday with 1240 citations. A single such paper in a 555-paper bin shifts the mean by several citations while leaving the median unchanged; the rank-based tests and medians are insensitive to these outliers.

\subsection{Announcement day}

\begin{figure}[ht]
\centering
\includegraphics[width=\textwidth]{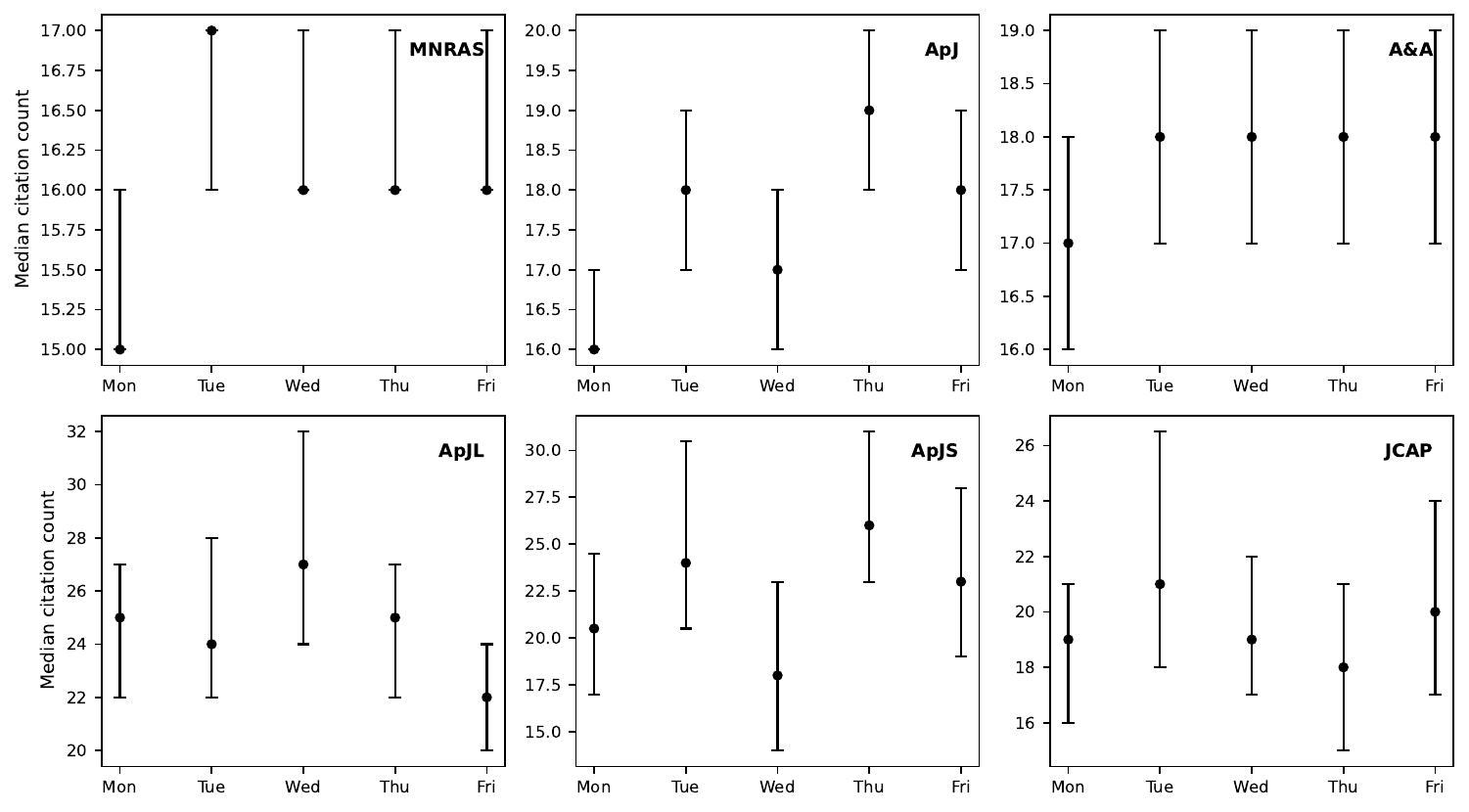}
\caption{Median citation counts with 95\% bootstrap confidence intervals as a function of the day the paper appears on arXiv, for each of the six journals.}
\label{fig:median_days}
\end{figure}

Table~\ref{tab:anndays} shows the  median citations by announcement day for all six journals, and Fig.~\ref{fig:median_days} displays the same information with bootstrap confidence intervals. The Monday listing --- which by construction absorbs all weekend submissions --- has the lowest median citation count in the three largest samples: MNRAS (15 versus 16--17 on other days; Kruskal--Wallis $p = 0.007$), ApJ (16 versus 17--19; $p = 5.6\times10^{-4}$), and, more weakly, A\&A (17 versus 18; $p = 0.018$). In the three smaller journals the lowest-median day varies (Friday for ApJL, Wednesday for ApJS, Thursday for JCAP) and the five-bin test is not significant for ApJL and JCAP, consistent with sampling fluctuations in bins of a few hundred papers.

\begin{table}[ht]
\centering
\caption{Median citations by announcement day (the day the paper appears on arXiv), with the Kruskal--Wallis $p$-value across the five days for each journal. The lowest median in each row is shown in bold.}
\label{tab:anndays}
\begin{tabular}{|l|r|r|r|r|r|l|}
\toprule
\hline
\textbf{Journal} & \textbf{Mon} & \textbf{Tue} & \textbf{Wed} & \textbf{Thu} & \textbf{Fri} & \textbf{KW $p$} \\
\midrule
\hline
MNRAS & \textbf{15.0} & 17.0 & 16.0 & 16.0 & 16.0 & $0.0071$ \\
ApJ   & \textbf{16.0} & 18.0 & 17.0 & 19.0 & 18.0 & $5.6\times10^{-4}$ \\
A\&A  & \textbf{17.0} & 18.0 & 18.0 & 18.0 & 18.0 & $0.018$ \\
ApJL  & 25.0 & 24.0 & 27.0 & 25.0 & \textbf{22.0} & $0.057$ \\
ApJS  & 20.5 & 24.0 & \textbf{18.0} & 26.0 & 23.0 & $0.012$ \\
JCAP  & 19.0 & 21.0 & 19.0 & \textbf{18.0} & 20.0 & $0.42$ \\
\hline
\bottomrule
\end{tabular}
\end{table}

We stress that the announcement-day and submission-day results are not independent findings: the Monday deficit is, to a large extent, the weekend-submission deficit re-expressed, because weekend submissions constitute a substantial fraction of Monday's listing. The robust, well-powered statement supported by these data is the weekend-versus-weekday contrast of Table~\ref{tab:weekend}; the Monday pattern is its visible consequence in the announcement-day view, resolvable only in the largest samples.

\subsection{Listing size}

Within our journal samples, Monday announcement batches contain the most papers per day in every journal (e.g., 12.6 versus 9.8 papers per day for Monday versus Friday in MNRAS; 8.2 versus 5.5 in A\&A), as expected since the Monday batch spans three days of submissions. This is consistent with the visibility-dilution mechanism of ref.~\cite{Dietrich2008}, in which papers in longer listings compete for reader attention, though we caution that our counts cover only each journal's papers rather than the full astro-ph listing and are therefore only a proxy.

\section{Discussion}
\label{sec:discussion}

Our results extend to astronomy the pattern reported by ~\cite{Delorme2025} for particle and nuclear physics: citation outcomes are flat across weekday submissions but lower for weekend submissions. The consistency of the effect across six journals with different sizes, publishers, and author communities and now across two broad fields of physics suggests a robust empirical regularity rather than an artifact of any single sample.

Two classes of mechanisms could produce this pattern, and the present data cannot distinguish them, which we enumerate below:
\begin{enumerate}
\item \textbf{Composition.} Papers submitted on weekends may differ systematically from weekday submissions in terms of author seniority, geographic distribution, time zones, paper type, or  circumstances of submission in ways that independently correlate with citation outcomes.
\item \textbf{Visibility.} All weekend submissions appear in Monday's enlarged listing, where each paper receives a smaller share of reader attention, and positional effects within listings are known to leave long-term citation traces~\cite{Dietrich2008,HaqueGinsparg2009,HaqueGinsparg2010}.
\end{enumerate}

A discriminating test available in principle is to compare, \emph{within} Monday's listing, papers submitted on Friday afternoon (after the 14:00~ET deadline) with papers submitted on Saturday or Sunday: both groups share the same listing, so a citation difference between them would point to composition rather than visibility. We defer this analysis to future work.
% TODO: or include it in this paper if Prof. Desai prefers -- the pipeline
% supports it with a small extension.

We emphasize the correlational nature of these results. They do not establish that posting on a weekend \emph{causes} lower citations, and they do not by themselves support submission-timing advice to authors: an association observed across papers does not predict the effect of an individual author changing their behavior.

The analysis comes with several caveats: differential arXiv coverage across journals (Table~\ref{tab:funnel}) imposes journal-specific selection; US-holiday schedule shifts mis-assign a small number of papers; citation counts were retrieved at a single epoch, so papers published in different years have had different amounts of time to accumulate citations (our year-stratified checks indicate the weekend pattern is not driven by any single year); and true listing positions cannot be reconstructed from ADS metadata alone. We also note that  sometimes citations to papers are sometimes incorrectly attributed~\cite{Will2014}.

\section{Conclusions}
\label{sec:conclusions}

Analyzing 37{,}173 refereed astro-ph papers published during 2020--2023 in six astronomy journals, we find that papers submitted to arXiv on weekends and consequently  appearing in Monday's enlarged listing --- receive systematically fewer citations than weekday submissions, with median deficits of 12--26 per cent, significant in five of six journals and directionally consistent in the sixth. Median citations are indistinguishable across the five weekdays. These findings extend to  astronomy the weekend citation deficit recently reported in particle and nuclear physics, and add to the evidence that visibility mechanics on arXiv are associated with long-term citation outcomes. Distinguishing compositional from visibility-based explanations is a natural next step.

\section*{Acknowledgments}

This research made use of NASA's Astrophysics Data System Bibliographic Services and of the arXiv API.

\section*{Data availability}
The analysis pipeline, cached data, and per-journal reports are available at 
\url{https://github.com/chirag003214/astroph-weekday-citations}.

% -------------------------------------------------------------
% References in EPJP format: numbered in order of appearance,
% initials-first author style, abbreviated journal names, one
% reference per number.
% -------------------------------------------------------------
\bibliography{name}
\iffalse

\fi 

\end{document}